\documentclass[12pt]{article}
\usepackage[utf8]{inputenc}
\usepackage{indentfirst}
\usepackage{graphicx}
\usepackage{biblatex}
\usepackage[ruled]{algorithm2e}
\usepackage{booktabs}
\usepackage{amsthm}
\usepackage{amsmath}
\usepackage{nomencl}
\makenomenclature

\title{Decentralized Federated Learning Based on Committees and Blockchain}
\author{ChaoQun Yang}
\date{May 2022}

\begin{document}

\maketitle

\begin{abstract}
Machine learning algorithms are undoubtedly one of the most popular algorithms in recent years, and neural networks have demonstrated unprecedented precision. In daily life, different communities may have different user characteristics, which also means that training a strong model requires the union of different communities, so the privacy issue needs to be solved urgently. Federated learning is a popular privacy solution, each community does not need to expose specific data, but only needs to upload sub-models to the coordination server to train more powerful models. However, federated learning also has some problems, such as the security and fairness of the coordination server. A proven solution to the problem is a decentralized implementation of federated learning. In this paper, we apply decentralized tools such as blockchain and consensus algorithms to design a support system that supports the decentralized operation of federated learning in an alliance environment, involving the exploration of incentives, security, fairness and other issues. Finally, we experimentally verify the performance of our system, the effect of federated learning, and the availability of privacy protection.

\end{abstract}

\newpage

\section{Introduction}
Currently, the accuracy of mainstream machine learning algorithms mainly relies on a large amount of data, from which machine learning algorithms extract features to predict or classify the next input. Therefore, the amount of data and data characteristics play a decisive role in the accuracy of machine learning algorithms, and the issue of data privacy protection is becoming more and more important. A major contradiction is that private data is private, and the key to improving the accuracy of machine learning algorithms may also lie in private data. Therefore, some privacy solutions have been proposed, such as a trusted execution environment, secure multi-party computation, federated learning, etc.

Federated learning is a proven solution to the privacy problem of machine learning. Each node does not need to directly expose the private data but uses its private data to train a sub-model locally and upload the sub-model to the coordination server. After the coordination server receives these sub-models, it aggregates the sub-models into a global model through an aggregation algorithm. The global model formed by the aggregation algorithm is generally slightly less accurate than training the model directly using data, but it largely solves the problem of privacy leakage.

The operation of the federated learning system completely depends on the coordination server, so the federated learning is still a centralized system. This will bring about a serious crisis of confidence.

1. \textbf{Federated machines suffer from insufficient incentives for data contributions.} The centralized coordination server is not transparent during operation, and the opaque data processing process makes it impossible for the participants to supervise the coordination server, resulting in a crisis of trust. This directly leads to the inability to fully expand the incentive mechanism.

2. \textbf{There are security concerns with federated machine learning.} A real problem is coordinating server precautions against malicious nodes. A possible effective way is for the participants to provide a certain pledge because this will increase the cost of malicious nodes. Implementing this mechanism requires the provider of the coordination server to master the pledge provided by the participating nodes. After the server provider masters the pledges of participating nodes, the fairness and security issues he faces also make it impossible to carry out relevant measures. The resulting crisis of confidence makes this mechanism difficult to implement.

Both of the above problems essentially point to a trust problem. In other words, if the participating nodes can fully trust the server provider, the problem will be solved. But the reality is not so.

It is very difficult to deal with the above problems on a centralized system. Therefore, our approach to solving these problems focuses on decentralized federated learning. The main goal of decentralized federated learning is to solve the trust problem caused by the centralization of coordinating servers. The key to solving this problem is to design a series of mechanisms to decompose the functions of the central node into multiple nodes. Multiple nodes will first face the problem of consensus, that is, how multiple nodes reach a consensus. Secondly, we need to design a set of data processing logic based on the consensus mechanism, so that the decentralized federated learning looks like the centralized federated learning in the way of working. Finally, based on the realization of the first two points, we have been able to design a simple decentralized federated learning.

However, it is far from enough to realize decentralized federal learning. We need to launch various mechanisms on its basis to ensure the normal operation of decentralized federal learning system, which will involve the design of incentive mechanism, punishment mechanism, security mechanism, privacy protection mechanism and other mechanisms. For example, in order to solve the problem that the enthusiasm of participating in the centralized federal learning system is not high, we need some incentive measures to encourage nodes to participate in the decentralized federal learning system, so that the centralized federal learning system will show good performance. In order to solve the problem of malicious nodes, we need a penalty measure to increase the cost of destroying the system. We also need a security mechanism to be responsible for the security of the whole operation process. At the same time, the privacy protection of participants is also an indispensable part.

\section{Related Work}
Recently, the wide application of big data and machine learning algorithms has made people pay more and more attention to the importance of privacy protection. In 2016, the federated learning proposed by Google\cite{mcmahan2017communication} can obtain global models by aggregating the models trained locally by users, and then obtain a more powerful model. Sub-models aggregation are usually done using FedAvg with FedSGD\cite{mcmahan2017communication}. The Mcmahan team's research pointed out that FedSGD is a special form of FedAvg, and the two are almost equivalent when the training volume is large enough, but FedAvg is significantly better than FedSGD in terms of communication times. Federated learning relies on the coordination server to aggregate the models of each participating node and evaluate the contribution of each participating node.

In fact, federated learning is still a centralized model training method. As the main participant of federated learning, the central node will have the possibility of a single point of failure. We can certainly avoid this problem by developing robust programs with powerful servers\cite{kairouz2021advances}. Obviously, this ability is very scarce. Centralized federated learning also has the problem of trust in the central node. It seems that only large companies have the ability to guarantee their actions and provide credibility guarantees. The reality is that small and medium-sized groups are eager to break out of the data silos while struggling to pay the cost of trust. Therefore, decentralized federated learning came into being. Decentralized federated learning is a system in which all nodes have equal status, unified responsibilities and obligations, and strict operating rules. Since Bitcoin \cite{nakamoto2008bitcoin} was issued at a price of \$0.0025 per piece in 2008, and it has skyrocketed to \$28,410 per piece in 2021, it has fully proved the feasibility of a decentralized system through practice. Ethereum \cite{buterin2014ethereum} is another decentralized transaction system based on blockchain architecture inspired by Bitcoin, its biggest feature is smart contracts. Smart contracts are DAPP development tools based on Ethereum. Using smart contracts can quickly build DAPP without the need for a decentralized design behind them. The success of these decentralized systems increases our confidence in implementing a decentralized federated learning system.

At present, there is also related research and implementation of decentralized federated learning.We divided their ideas into two categories.

1. \textbf{Architecture using peer-to-peer network or blockchain.} Braintorrent\cite{roy2019braintorrent} is a completely peer-to-peer decentralized federated learning system. Braintorrent only discussed the decentralization of federated learning and the operation process after decentralization but did not involve issues such as incentives and penalties. In fact, a completely peer-to-peer system that does not use blockchain guarantees speed, but it is difficult to implement mechanisms such as incentives and penalties, because the problem of staking is difficult to solve, and blockchain is just a good choice to solve the problem of staking. DeepChain\cite{weng2019deepchain} is a decentralized federated learning system with a relatively complete mechanism. DeepChain chose to create its own blockchain and issue new coins to complete the incentive and punishment mechanism, and used the committee mechanism to reduce the pressure on the system. Specifically, the committee is responsible for packaging transactions, which is a kind of miner in the form. They choose a leader through a consensus mechanism. The leader is responsible for packaging various transactions, and other committee nodes are responsible for verifying the correctness of the block. After that, the leader will obtain the system reward. This way of realizing incentives and punishments by issuing new coins essentially makes DeepChain a closed model market. In addition, DeepChain uses homomorphic encryption\cite{gentry2009fully} to ensure the security of the model, but the use of homomorphic encryption also means that it is difficult for the committee to obtain the quality of the model, which makes the price of the model difficult to measure, and also makes poisoning attacks are hard to defend\cite{lyu2020threats}. FLChain\cite{majeed2019flchain} has conducted a detailed discussion on incentive issues on the basis of DeepChain and has more diverse and reasonable incentive mechanisms. The Youyang Qu team has done a good job on poisoning attacks\cite{qu2020blockchained}. For the poisoning problem, they try to analyze the poisoning process dynamically using Nash equilibrium. Their results show that it is difficult to carry out a poisoning attack in a robust decentralized system, and for the initial stage of a centralized system when the overall computing power of the system is not strong, the overall impact of a successful poisoning attack on the system is relatively small. big. The team of Pokhrel et al.\cite{pokhrel2020decentralized} applied decentralized federated learning to the field of autonomous driving. BLADE-FL\cite{li2021blockchain} uses differential privacy technology to protect participating nodes, which is mainly used to defend against model inversion attacks. Their team also proposed the concept of lazy nodes and analyzed differential privacy and the impact of lazy nodes on accuracy. A lot of experiments have been done. Their experiments show that the increase in the proportion of lazy nodes makes the overall accuracy drop significantly. Qi's team \cite{qi2021privacy} pointed out through experiments that differential privacy can effectively defend against model flipping attacks, which needs to be found in terms of data availability and privacy protection. a balance. The architecture of IPLS\cite{pappas2021ipls} is completely based on IPFS. Every piece of information in IPLS can be regarded as an IPFS file and obtained consensus in IPFS. IPLS does not involve any incentive or punishment mechanism but is only an innovation of the architecture.

2. \textbf{Architecture using smart contracts as controllers.}BAFFLE\cite{ramanan2020baffle} uses smart contracts as aggregators to implement the aggregation function in federated learning. Using smart contracts as controllers can make the system simple, and the system is very robust. As far as we know, each block of Ethereum stores about 70 transactions, and a block is dug every 13 seconds or so, we can simply calculate that Ethereum can process about 5-6 transactions per second. A smart contract call is considered a transaction in Ethereum, which means that using a smart contract as an aggregator makes the system inefficient.FedBC\cite{wu2020fedbc} adopts a ring-shaped structure and cooperates with smart contract scheduling to protect the privacy of participating nodes. The model is aggregated backward one by one, and the current node only knows the overall aggregation model passed by the previous node.

\section{The Proposed Architecture}

\subsection{System Architecture}
According to our research on related papers, we conclude that decentralized federated learning can be roughly divided into two types, namely blockchain architecture and smart contract architecture. The advantage of using blockchain architecture is that the system design is flexible and changeable, and designers can design different decentralized systems according to different business scenarios. However, the flexible system design will inevitably bring some troubles, the most fatal disadvantage of which is the poor incentive effect. The reason for the poor incentive effect is that the value of the private chain is low, and the private chain with fewer participating nodes does not have the strong currency properties of the public chain. If the smart contract (Mainnet) architecture is used, the incentive mechanism problem of the blockchain architecture can be solved. We also mentioned before that smart contract calls are not fast, so this will seriously slow down the system. Our solution is to do off-chain consensus first and then on-chain consensus. This eliminates the need for smart contracts to fully bear the pressure on the system. Most of the pressure of the system will be borne by off-chain nodes, and the smart contract only accepts and verifies the consensus results of off-chain nodes and is responsible for the rewards and punishments of the system. This solution not only improves the speed of system operation but also effectively solves the incentive problem generated by the blockchain architecture. In the selection of the consensus algorithm for off-chain consensus, we chose the PBFT algorithm. The advantage of this algorithm is that the consensus efficiency is high and energy saving. The downside is that its performance drops sharply as the number of participating nodes in the system rises. To solve the performance problem of the PBFT algorithm, we introduced a committee mechanism, that is, according to the contribution of the participating nodes in the system, some nodes are selected as the committee, and the PBFT algorithm is run between the committees to increase the maximum number of participating nodes in the system. Finally, considering that the sub-models submitted during the federated learning process may also expose private information, we use differential privacy technology to process the sub-models to further protect user privacy data. In summary, we propose our system architecture, which we call DFLBCB. The architecture of DFLBCB is shown in the following figure.

\centerline{\includegraphics[scale=0.45]{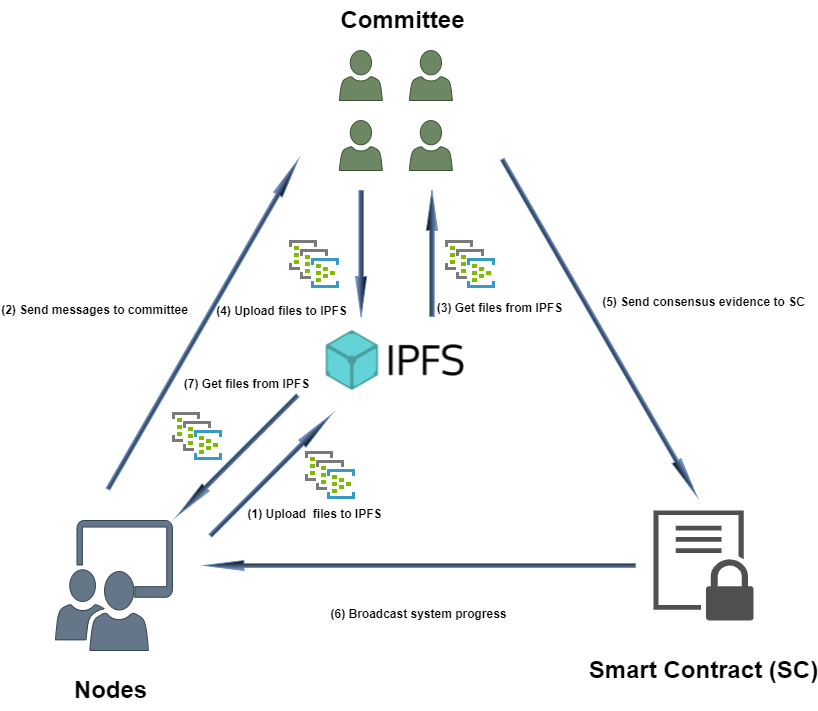}}

\subsection{Offline Consensus}
Similar to other decentralized systems, the system is less robust at an earlier stage of operation. The reason is that there are fewer participants in the early system, and the system as a whole is not active enough. Malicious nodes can enter the key positions of the system at a low cost, and then destroy or control the entire system. This is also the reason why various attacks are easier to carry out in the early stage of the decentralized system\cite{sayeed2019assessing}\cite{ye2018analysis}. Therefore, DFLBCB relies more on offline consensus during the initialization period. Offline consensus means that the initialized nodes, the value-accuracy curve, and the deployment address of smart contracts are all agreed upon in advance by the project participants offline, which also makes the system controllable and safe in the early stage of operation. So we need to prepare:

1. Support nodes. Logically, we generally refer to the nodes that support DFLBCB running as support nodes. A support node is just a logical concept and not a physical concept. This means that you can deploy a support node on any machine, even though multiple nodes are already deployed on that machine. A support node must run at least the IPFS client, and Geth client, and connect to other support nodes. 

2. State contracts. A state contract is a smart contract written by us. The state contract maintains many states, which represent the overall state of the current DFLBCB, so we call it the state contract in DFLBCB. The status of DFLBCB includes many factors, such as the current iteration round, node list, committee information, global model ID, global model score, current stage, pledge list, and a series of information reflecting the current DFLBCB status. Maintaining this information is to coordinate the operation of DFLBCB important way. In addition, the state contract not only records the current state but also records the past state for query. 

3. Registration and pledge deposit. To prevent DFLBCB from repeatedly calling the state contract due to pledge during each iteration, if users want to participate in DFLBCB, they need to pledge to the state contract in advance. This process is called registration. The benefit of registration is to improve the performance of DFLBCB. As can be seen from the previous discussion, the efficiency of calling smart contracts is not high, so we should avoid calling smart contracts as much as possible. After registration, the state contract will create a fund pool for the node in advance. Whenever the node needs to pledge, it does not need to call the state contract every time. Instead, collect the nodes that need to be pledged and then call the state contract. The state contract is unified from Debit from the fund pool.

4. Accuracy-price curve. The accuracy rate-price curve is depicted by multiple two-dimensional coordinates (x, y). The horizontal axis represents the accuracy rate, and the vertical axis represents the corresponding model price under the current accuracy rate. Every time a global model is generated, the state contract will record the accuracy of the model and find its corresponding price according to the accuracy-price curve. The search principle follows the average principle, which means that the price of the model is at the average value of the price corresponding to the adjacent accuracy rate. Therefore, more scatter points mean a more precise accuracy-price curve.

5. Training data and model parameters. Training data is local private data of participating nodes, which is the value of federated learning. Decentralized federated learning inherits the advantages of traditional federated learning, that is, the training data does not leave the local area, which largely protects the privacy of participating nodes. Similar to traditional federated learning, in decentralized federated learning, participating nodes need to prepare training data in advance to train sub-models locally. The size of the training data should also be set in a range. Excessive data size may lead to a long training time for the model, which may cause the model trained in this round to not be submitted within the specified time. It should be noted that to improve the operation efficiency, DFLBCB will set a maximum completion time based on the completion time of most participating nodes at each stage. Participating nodes that exceed the maximum completion time will not be able to continue to submit, and can only wait until the next round. Submit again, and the resulting consequences will be borne by the participating nodes. It is worth noting that in the pre-preparation stage, the participating nodes should explain the structure and parameters of the model in as detail as possible. Unclaimed uploading of unknown models is considered malicious and penalized. Furthermore, the work of Bob et al. [16] showed that if all nodes are trained based on a consistent initial global model, then we will achieve higher accuracy. Therefore, the initiator of the project needs to prepare an initialized global model in advance, and the participating nodes will train based on the global model

6. As high-performance machines as possible. The consensus information of DFLBCB is mainly undertaken by the committee, and the committee nodes will bear more processing pressure than ordinary participating nodes. Committee nodes with different performances have different transaction processing performances. If most of the committee nodes are high-performance devices, and an ordinary PC is elected to the next committee at this time, it is a very bad thing, because this computer The PC's ability to process transactions is much lower than other devices, and the time to reach a consensus will be significantly longer than other devices. In the design of DFLBCB, the minimum requirement for consensus is the consensus of the majority of people rather than the consensus of all, in which case DFLBCB will not wait for poor-performing devices. To ensure the security of DFLBCB, we have strict requirements for committee nodes. If there is any non-response situation on the committee node, that is, if the specified thing is not completed within the specified time, DFLBCB will punish the committee node regardless of the reason.

7. Start DFLBCB. DFLBCB will be started by a transaction, which is to call the initialization function in the state contract, which is generally called by the project initiator. The initialization function needs to pass in many parameters, such as initial node information, accuracy-price curve, initial global model, privacy parameters, and a series of parameters that have been agreed upon in advance.

\subsection{System Workflows}
The system workflows of DFLBCB are shown in the figure below. We divide an iteration of workflows into 4 stages, namely Elect, Pledge, Commit, and Work. These four stages complete different tasks respectively, and this round of iteration is completed when the four stages are completed. Below we describe these four stages in detail.

\centerline{\includegraphics[scale=0.45]{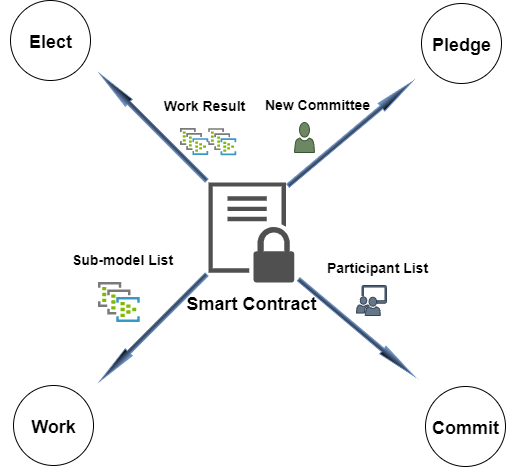}}

\subsubsection{Elect Phase}

The purpose of the election is to elect the nodes in DFLBCB who have contributed greatly and are willing to provide pledges to process system affairs. In DFLBCB, the most intuitive way to measure the contribution of a node is to count the accuracy of its submitted models in the past. Higher model accuracy means more data input. However, it is difficult to penalize committee nodes only by relying on the accuracy of the model. Therefore, DFLBCB also requires committee nodes to submit pledge deposits. There is a minimum value for the submission of the pledged deposit, which is agreed upon in advance in the pre-preparation stage. This also means that the higher the pledged deposit, the more right to be elected as a committee node. However, the introduction of pledge deposits not only strengthens trust but also weakens the degree of closure of the system, making the system vulnerable to external shocks. This is the disadvantage of introducing pledged deposits. An attack method is implemented as follows: the attacker registers a large number of accounts with sufficient funds and controls these accounts to participate in elections and invests a high deposit, and the attacker controls the committee and then retrieves the funds, thus achieving a cost-free attack. To solve this problem, instead of adding the pledged deposit and the accuracy rate after equivalent conversion, we assign a higher weight to the accuracy rate. So we propose the following algorithm to help DFLBCB conduct elections. The weights of participating nodes are as follows:

$$w = p*eth+(1-p)*acc*\bar{eth} $$ 

Among them, $p$ is the weight adjustment coefficient, $\bar{eth}$ is the average pledge number of the top $Need$ nodes, and $acc$ is the average accuracy rate
    \begin{algorithm}[H]
    \caption{Election algorithm}
    \KwIn{Candidate List:$ CList $ ; Number of candidates:$Need$}
    \KwOut{electee information:$Result$}
    $//$If there are not enough candidates, return directly\;
    \If{$ len(CList) < Need$ }{\Return $Result$}
    \For{$candidate$ $in$ $CList$}{
    Calculate the weight w of each candidate\;
    }
    $//$Sort candidates by weight w\;
    $Result \gets Sort(CList)\ by\ w$\;
    $//$Take the first Need candidates as the result\;
    $Result \gets \ Result[:Need]$\;
    \Return $Result$\;
    \end{algorithm}

The algorithm shows that all nodes eligible to be elected to the committee will share a certain pledge deposit base.

1. For participating nodes that provide high deposit and high accuracy, the node ranks high.

2. For nodes that provide a high deposit and a low accuracy rate, the ranking of the node is uncertain, but it will increase the success probability of a node with a small deposit and a high accuracy rate. The node may also be elected if the node provides enough collateral.

3. For participating nodes that provide less deposit and high accuracy, the node ranks higher.

4. For participating nodes that provide less deposit and lower accuracy, the node ranks at the bottom.

Scenario 1: Assuming that the total number of registrants is N and the election ratio is P, then NP committee nodes will be selected. According to the description of the algorithm, DFLBCB will select the top NP as committee nodes. Once the committee nodes are elected, they will be recorded in the state contract in order. 

Scenario 2: Each candidate node sends the candidate information to the master node, and after receiving a certain number of responses, the candidate node can confirm that the committee has received the candidate information. 

Scenario 3: Before executing the election algorithm, each current committee member is guaranteed to know the global election information. Only if each committee node knows the global election information, the committee can choose the same result according to the algorithm 1. 

Scenario 4: After executing the election algorithm, each committee node packs the information of the participating nodes and takes the hash value \cite{dworkin2015sha} to upload the status contract. The state contract decides whether it can enter the next state - the pledge phase.

\subsubsection{Pledge Stage}
At this stage, there may be a committee turnover, i.e. new committee members replace old committee members. This stage will also decide which nodes will participate in this round of training. The model training method of DFLBCB is iterative, and the participating nodes decide whether to participate in this round of iteration. The staking phase is designed to count how many nodes will participate in this round of training. Because confirming participation is a short-lived process, submitting a model is a time-consuming process. Therefore, similar to the election phase and the pledge phase, a fixed-time timer can be used to solve the state transition problem. 

Scenario 5: If this stage is transferred from the election stage, the committee will be re-elected first. The change information is stored in the event information of the smart contract. 

Scenario 6: If you participate in this round of iteration, on the premise of ensuring your capital pool is sufficient, you can send participation request information to the committee master node, and confirm that the information has been accepted by the committee after receiving a certain number of committee responses. 

Scenario 7: The committee generally considers all staked nodes to be legitimate nodes. In other words, the committee only performs primary verification, that is, to verify that the node is registered and that the signature \cite{miller1985use} matches, without verifying the account balance. 

Scenario 8: The committee packs the participating nodes, takes a hash value, and uploads the status contract. The state contract decides whether to move to the next state - the commit phase.

\subsubsection{Commit Phase}
At this stage, the state contract will announce the nodes that participate in the pledge stage and meet the conditions through events. The committee node will gather the sub-models provided by the participating nodes at this stage. 

Scenario 9: If the participating node exists in the list of nodes that meet the conditions announced by the state contract, then the participating node needs to submit the sub-model at this stage, submit the request information, and confirm that the information has been received after receiving a certain number of committee responses. The committee accepts. 

Scenario 10: The training time of the sub-model may be very long, so the submission process may take some time, and DFLBCB will continuously query the training status of the sub-model. When the model has not been submitted within the longest training time, DFLBCB will terminate the query process and give up the current round of submission. The maximum training time of the model is obtained according to the algorithm 2. Algorithm 2 will make DFLBCB accept as many sub-models as possible on the premise of collecting most of the sub-models. 

Scenario 11: The committee packs the participating nodes and takes a hash value and uploads the status contract. The state contract decides whether it is possible to enter the next state - the work phase.

\subsubsection{Work Phase}

At this stage, the state contract will publish the sub-models trained by each participating node through events. The committee nodes will evaluate and merge the sub-models submitted by each node at this stage. The evaluation method is based on the pre-consensus test set.The test accuracy of the sub-model is accurate to three decimal places and then multiplied by 100 to get a three-digit integer. After the test, the sub-models are aggregated into a global model, and the accuracy of the global model is tested again. 

Scenario 12: Submit the results of this work to other nodes of the committee and submit request information. Confirm that the information has been accepted by the committee after receiving a certain number of committee responses. 

Scenario 13: The work time may be long, so the submission process may take some time, and DFLBCB will continuously query the work status. When the model has not been submitted within the longest working time, DFLBCB will terminate the query process and abandon the current round of submission. The maximum working time of the model is derived according to the algorithm 2.

Scenario 14: The committee packs the participating nodes, takes a hash value, and uploads the status contract. At this time, the committee has two possible state transitions. One is to move to the election phase, and the other is to move to the staking phase. The difference between the two is whether the next iteration requires re-election of the committee. Overall, DFLBCB has a total of 4 stages, each stage takes a certain amount of time, and each stage saved will greatly improve the performance of DFLBCB. In addition, it seems unnecessary for DFLBCB to select a committee node for each iteration. After all, the design significance of committee nodes is to maintain system operation and improve consensus efficiency. But only one committee election is not conducive to the robustness of the system. DFLBCB's approach is to agree on a parameter $ElectTimes$ in advance, which indicates how many iterations the committee can last for each election round. Whenever the number of iterations of the committee exceeds $ElectTimes$, DFLBCB will enter the election phase, otherwise, it will enter the staking phase.

\begin{algorithm}[H]
    \caption{State Transition Algorithm}
    \KwIn{maximum acceptable time:$mt$;request arrival time:$arrs$;number of requests:$tot$}
    $//$ This algorithm is used to determine whether a state transition has occurred\; 
    start timer$1$,set time to $mt$\;
    $//$ Geting the average number of request arrival times\;
    $t=\bar{arrs}$\;
    \If{$arrs/tot>=80\%$}{
    start timer$2$,set time to $t$\;
    }
    \For{$TRUE$}{
    \If{receive new requests}{
    reset timer $2$\;
    }
    \If{trigger timer $2$}{
         $t=t/2$\;
         \uIf{$t<=0$}{
         End of this stage\;
         }
         \uElse{
         Start the timer $2$, set the time to $t$\;
         }
    }
    }
\end{algorithm}
\subsection{Summary}
In general, each stage can be decomposed into result inheritance, message consensus, state confirmation, and advancement. The correct operation of the scenarios 5, 10, 13, requires listening to the relevant events of the state contract and integrating the messages. In the scenario 2, 6, 9, 12, each committee node is mostly processing consensus messages and progressing through the algorithm 2 in the continuous calculation phase, because PBFT is a strong consistency state machine replica replication algorithm, algorithm 2 is also a deterministic algorithm, so we can always happen the transition of DFLBCB state in a deterministic state. Scenarios 4, 8, 11, 14, describe the process of committing to a state contract. The state contract first verifies the first contract call that submitted no less than F+1 confirmation messages. After the state contract verifies the contract call, the state contract can judge that the message has been agreed upon by the committee nodes. The state contract then verifies the legitimacy of each specific request in detail. For example, in scenario 6, the participating nodes include at least $<it, status, money, address, sig> $ in the request information, which represents the current round, the state, the pledge amount, and the signature, the participating node's address. Through this information, the state contract can judge whether the request is legal according to its state, and deduct the fee if it is legal. Without $<it, status>$ to uniquely identify when the request was made, the leaked request could be exploited maliciously. It can be seen that the establishment of the stage state can also effectively deal with replay attacks.

\section{Consensus Mechanism}
\subsection{Analysis Of Consensus Problems In DFLBCB}

In the previous section, we roughly described the operation of DFLBCB. By summarizing the scenarios of DFLBCB during operation, we divide the consensus problems in DFLBCB into 2 categories: 

1. Willingness-driven. A willing-driven consensus request is a consensus request initiated by participating nodes at the right time according to the principle of voluntariness. The willingness of participating nodes drives the generation of such requests. For example scenarios 2, 6, 10. In these scenarios, whether a participating node sends a request only depends on the current global state of DFLBCB and its willingness to participate. For this type of consensus problem, as long as the committee nodes can reach a final consensus state, it is sufficient. In scenarios 2, 6, 10, it is assumed that 3 participating nodes send requests M1, M2, and M3 respectively. All committee nodes receive all requests in any order to achieve a consistent result. In summary, for the will-driven consensus problem, there is no need to pay attention to the consensus process, but only to achieve a final consensus. 

2. State-driven. A state-driven consensus request is a consensus request that is controlled by an algorithm and generated based on the current consensus state of committee nodes. Conceptually, new states are derived when a committee reaches a certain state, such as the scenario 4, 8, 12, 15. In these scenarios, committee nodes face a problem - how to choose the right time to end the current phase. A naive idea is that we can end the current phase and move on to the next phase when we agree on all the willingness-driven requests. Unfortunately, DFLBCB cannot predict in advance - DFLBCB cannot confirm the specific number of participating nodes in advance, and even if DFLBCB can know how many participating nodes are in advance, it is impossible to wait for all nodes to submit requests indefinitely. At most, DFLBCB can only infer how many participating nodes are in this round based on historical information. Therefore, we design an algorithm 2 to decide when to end the current state.

The above two types of consensus problems combined with the operating environment of DFLBCB determine what consensus algorithm we will use. Willing-driven consensus requests determine that we will use any eventual consensus algorithm. The state-driven consensus request determines that we must adopt a strong consensus algorithm because only a strong consensus algorithm can guarantee the consistency of the previous state. DFLBCB operates in a less malicious alliance environment, which determines that the consensus algorithm we adopt must be fault-tolerant. At the same time, the alliance environment with a low malicious degree shows that we will not adopt the POW\cite{mingxiao2017review} algorithm used in high malicious degree. To sum up, we finally realized the consensus algorithm PBFT-SC based on PBFT\cite{castro1999practical} consensus algorithm and smart contract\cite{buterin2014ethereum}.

    \begin{algorithm}[H]
    \caption{PBFT-SC}
    \KwIn{Received consensus request:$M_{1},M_{2}...M_{n}$}
    $//$Committee nodes will generate a response $R_{n}$ for each message $M_{n}$ agreed upon by the committee nodes\;
    $R_{1},R_{2}...R_{n}$=$PBFT$($M_{1},M_{2}...M_{n}$)
    $//$Suppose now that the committee node $m$ sends a state transition consensus $M_{end}$ according to the algorithm 2\;
    $//$When node $m$ receives $>=F+1$ responses to $M_{end}$ requests, it can submit it to the state contract\;
    $R_{end_{1}},R_{end_{2}},R_{end_{3}}...R_{end_{F+1}}$\;
    The status contract checks the signature and status of $F+1$ responses\;
    The status contract performs signature and status checks on requests from participating nodes\;
    The state contract checks the participating nodes according to the node information maintained by itself\;
    All checks are completed, the state contract moves to the next state, and an event notification is issued\;
    \end{algorithm}
 
DFLBCB continuously judges the best commit point by running the algorithm 2. When DFLBCB detects that the current state has reached the commit point, it runs the PBFT-SC algorithm. The PBFT-SC algorithm is divided into 2 stages: 
 
PBFT stage. Each DFLBCB node that detects a commit point sends a consensus request End<address,type,hash,sig>. The four parameters represent the address of the committee node, the current stage type, the resulting hash, and the signature. The resulting hash is the process of calculating the result of the consensus information before all the submission points and calculating the hash value of the result. For example, for a scenario 3, the committee node runs the algorithm 1 and the result is the addresses of the first NP committee nodes. Send the End<address,type,hash,sig> request to all node committee nodes for consensus. For each committee node, when the End request reaches the PBFT-Reply stage, it starts to check whether the local result hash value is the same as the resulting hash provided by the End request. If they are the same, return Reply<address,type,hash,sig,result>. The first four parameters of Reply have the same meanings as the four parameters of End. The difference is that address and sig of End represent the sender's address and signature, Reply represents the address and signature of the responder, and the result represents the result. According to the description of the PBFT algorithm, when the sender gets responses from F+1 different nodes, it can be considered that the message has been consensus, and these responses will be used as proof of state transition. Note that several committee nodes may arrive at the commit point at the same time, and they may all hold state transition proofs.

SmartContract stage. Scenario 16 for every committee node that holds greater or equal to F+1 state transition proofs, they have the right to push DFLBCB state transitions—they can all call state contracts to submit state transition proofs. A state contract can only accept the first legal proof of state transition. The state contract verifies the legitimacy of the state transition evidence through the algorithm 3. For each state transition attempt, the state contract requires the participating nodes' original set of requests and proof of state transition. The state transition evidence is used to prove that at least $F+1$ nodes have confirmed the state, that is, the state has reached the global consensus of the PBFT algorithm. The original request contains information such as request time, global state, maximum acceptable cost, signature, etc., and it is immutable to prove that it is indeed the participating node that authorizes the committee node to call the contract and deduct the fee. After completing the proof of the two, the state contract considers the result to be credible, then performs the corresponding operation according to the result, and then obtains a final result. The state contract will notify all listening nodes through events. Nodes listening to the event can make state changes immediately. At the same time, the event contains the final result of the previous stage, and the monitoring node then updates the local information according to the result information.
\subsection{Security Of Committee Node Switching}
After each election, DFLBCB will face the problem of node switching. Generally speaking, some of the old committee nodes are still in the A state, some of the new committee nodes are in the successor state B of the A state, and the PBFT-SC algorithm is split. In conclusion, the new committee node in state B no longer belongs to the old committee node, and the old committee node in state A still thinks that state B has not yet been generated. For honest nodes that are still in the A state of the old committee node and elected to the new committee member of the B state, we record them as PA nodes, which may happen as follows. 

1. The PA node still in the A state increases the proportion of malicious nodes in the A state because some former AP nodes have transitioned to the B state
2. The new committee node in state B increases the proportion of malicious nodes in state B because some PA nodes do not reach the new state B

For case 1. PA nodes may agree on some information under the deception of malicious nodes, but because at least F+1 non-Byzantine nodes have completed Reply, PA nodes can't receive 2F+1 confirmations at any stage, so It is impossible for the PA nodes to agree on any misinformation, but may remain stalled. Furthermore, no matter what state the PA node is in, how high the proportion of malicious nodes is because the PA node is not a Byzantine node, the PA node will eventually be corrected by the state change event of the state contract.

For case 2. For a new committee node that is switching to the B state, it may take a while to switch to the new node. But these switching nodes are not subjectively malicious. For consensus requests, these nodes will reject the request because they have not completed the switch, and will not actively send malicious requests. Therefore, in the worst case, the system will be in a state of hiatus for some time. Eventually, these nodes will gradually complete the update under the synchronization of the state change event of the state contract.

To sum up, if the committee switch meets the fault tolerance requirements of PBFT, DFLBCB will be temporarily suspended in the worst case, and then the state contract will refresh the state of the committee node through the state change event. Therefore, the PBFT-SC consensus algorithm can guarantee the security of committee switching.

\subsection{The Role Of Smart Contracts}
Record global information and notify state switching. If it is complicated to simply use the PBFT algorithm to implement the secure switching of states, a set of algorithms for the secure connection of new and old states should be designed, such as the above-mentioned committee switching problem. The new committee node must design a set of information transfer protocol with the old committee node to ensure that the new committee node can learn the composition of the new committee from some old committee nodes, and at the same time ensure that the old committee node can safely exit or state transfer. The smart contract can also provide the initial consensus function. The project party can directly write the initial consensus information into the smart contract, and notify each node of the project's initialization data through events, thus avoiding the problem of incorrect configuration during project initialization. The global historical state and state transition evidence recorded by the smart contract can also help us check the running records and fault information of DFLBCB.

Efficient pledge. Registering in DFLBCB generates a fund pool, whether it is election pledge or model submission pledge, whenever the state contract verifies the first legal state transfer evidence, the state contract will extract the node information participating in the pledge. Since the information is signed by the participating nodes, it must be proved that the information is sent by the participating nodes, and the state contract can deduct the fees for the nodes participating in the pledge in the first call, avoiding the time-consuming of a single deduction.

Security is scalable. Currently, DFLBCB runs in a less malicious alliance environment, and DFLBCB may run in a more malicious environment in the future. Different security should correspond to different environments, so in order to realize the scalable consensus security of DFLBCB, we have introduced smart contracts. In a consortium environment, we can run the ETH private chain, and in a more malicious environment, we can directly use the ETH public chain and increase the difficulty of the election.

\section{Rewards And Penalties}

\subsection{Common Behavior}

\resizebox{\textwidth}{!}{
\begin{tabular}{@{}ccc@{}}
\toprule
\multicolumn{1}{c}{Events}                          & Rewards            & Punishments              \\ \midrule
Committee correct aggregation model                 & +1                 & X                        \\
Committee evaluates model correctly                 & +1                 & X                        \\
Committee Submit State Contract Verification        & Algorithm 4  & X                        \\
Committee elected masternode and working            & Algorithm 5   & X                        \\
Committee Node Unanswered                           & X                  & Exponentially decreasing \\
Committee actively responds                         & +1                 & X                        \\
Model not submitted after participating node pledge & X                  & Linearly decreasing      \\
Participating nodes submit low quality models       & X                  & Exponentially decreasing \\
Participating nodes submit high-quality models      & Algorithms 5 & X                        \\
Participating nodes submit regular quality models   & Algorithms 6 & X                        \\ \bottomrule
\end{tabular}}
\subsection{Reward And Punishment Measures}
algorithm 4 believes that if the fee for a committee node to call the state contract is greater or equal to the reward that can be obtained in this round, then there is no need for the node to actively call the state contract. Therefore, the algorithm tries its best to make the cost of calling the state contract less than or equal to the reward that can be obtained in this round. In this iteration, the total cost of calling a state contract is $ts$, the current price of $gas$ is $gp$, the total prize pool in the system is $tm$, and $gp$ remains unchanged when each committee node calls the smart contract. At this time, there are $m$ nodes in total, the incentive score obtained is $p_{i}$, and the reward score obtained in this round is $s$.

\begin{proof}
$$ $$
If the node submits to the state contract:\\
$$\frac{p_{n}+s}{\sum_{n=1}^{m} p_{i} + s} * tm $$ \\
If the node does not commit to the state contract:\\
$$\frac{p_{n}}{\sum_{n=1}^{m} p_{i} + s} * tm $$\\
That is to prove:
$$ts>= \frac{p_{n}+s}{\sum_{n=1}^{m} p_{i} + s} * tm - \frac{p_{n}}{\sum_{n=1}^{m} p_{i} + s} * tm$$\\
$$\frac{tm*s}{\sum_{n=1}^{m} p_{i} + s} >=ts$$\\
$$\frac{1}{\sum_{n=1}^{m} p_{i} + s} <= \frac{1- \frac{ts}{tm}}{\sum_{n=1}^{m} p_{i}} $$\\
$$s>= \frac{ts*\sum_{n=1}^{m} p_{i}}{tm-ts} $$\\

\end{proof}

Algorithm 5 believes that the consensus pressure on the main committee node is far greater than that of the ordinary committee node. Therefore, such nodes should be deployed to more powerful computers, which also means more rewards. The way we evaluate system stress relies on the amount of consensus information. The more consensus information, the more consensus tasks the main committee node undertakes. When a committee node collects F+1 Reply with the same result and submits it to the state contract, after the state contract is verified, it can know the approximate number of consensus messages at this stage. According to our design principles, after one iteration, the pressure on the main committee node on consensus requests is about 2 times that of other committee nodes, and the main committee node will receive $n$ times the rewards generated by other positive responses.

$$ n = f(x)= 1+x/100$$
$n$ is a coefficient between (1,2], which is determined by the total number of consensus information at this stage. According to our experience, the (1,2] interval corresponds to all requests for the number of 100 nodes. For example, there are now 10 nodes sending out Request, $n=f(10)=1+10/100=1.1$, if there are 100 nodes making requests now, $n=f(100)=1+100/100=2$. The reason for this design is that the less information is requested in the system, the closer the pressure on the committee master node is to the pressure on other committee nodes, and the less reward.

Algorithm 6 believes that participating nodes submit models of general quality and high-quality models are beneficial to the operation of the system, especially the submission of high-quality models. Submitting low-quality models is not conducive to the operation of the system. Therefore, nodes that submit normal quality, high-quality models should be rewarded, and nodes that submit low-quality models should be penalized. Due to the nature of federated learning, the accuracy of the global model is generally higher than the local accuracy, and retraining the client based on the global model will lead to a decrease in the accuracy of the model.

Therefore, the system will reward the models with the highest quality, and discard the low-quality models generated by malicious nodes. We take the median of the model's accuracy ranking as the basic judgment value. Models higher than the basic judgment value will be rewarded. Participating nodes that are lower than the judgment value N\% will be regarded as malicious, and the pledged deposit will be deducted exponentially. At present, we have largely solved the situation that malicious nodes generate low-accuracy models to attack, but we still cannot solve the problem of lazy nodes and other attack methods.

Linear and exponential decrease of pledge deposit. Linearly decreasing means that the error will not cause too much damage to DFLBCB, but because the behavior is objectively malicious, the pledge is deducted in a linearly decreasing manner. Generally, 8\% of the maximum pledge amount is deducted each time. The exponential decline means that this behavior will cause serious damage to DFLBCB, but DFLBCB initially believes that it does not have subjective maliciousness, so the initial deduction of the pledged deposit is not large. However, with the increase in the number of malicious behaviors, DFLBCB has to think that the behavior is subjectively malicious. Therefore, for this kind of behavior, DFLBCB deducts a small amount of pledge deposit from the node when the behavior occurs, to tolerate its objective malicious behavior. As the malicious behavior continued, DFLBCB penalized the node massively. Exponential functions happen to have similar properties, so DFLBCB calls this method of deposit deduction an exponentially decreasing method. Generally, we specify the following exponential deduction function as our default function.
$$ f(x)=2^x $$
\begin{algorithm}[H]
\caption{Committee commits to state contracts}
\KwIn{Current incentive score for each node:$p_{1},p_{2},p_{3}...p_{m}$;The total amount of pledge deposit invested:$tm$}
$//$Get the current remaining GAS before each contract call.\;
$gas1=msg.gas()$\;
do something\;
$gas2=msg.gas()$\;
$//$After each contract call, the current remaining GAS is obtained, and the total cost of GAS is calculated.\;
$ts=gas2-gas1$\;
$//$After each contract call, the state contract is responsible for calculating the reward points in real time.\;
$s= \frac{ts*\sum_{n=1}^{m} p_{i}}{tm-ts} $\;
The status contract records to the corresponding account $+s$ reward points\;
\end{algorithm}

\begin{algorithm}[H]
\caption{The committee is elected as the masternode and works normally}
\KwIn{The number of all unique consensus requests at this stage:$RS$;The score that the committee master node should get at this stage:$PS$}
do something\;
$//$Calculate the reward coefficient of the master node
$s=(1+RS/100)*PS$\;
$//$After each contract call, the state contract is responsible for calculating the reward points in real time.\;
The status contract records to the corresponding account $+s$ reward points\;
\end{algorithm}

\begin{algorithm}[H]
\caption{Participating nodes submit model quality}
\KwIn{Scores of submodels sorted:$s_{1},s_{2},s_{2}...s_{m}$;Malicious rate:$n$}
$mids$ $\gets$ $s_{1},s_{2},s_{2}...s_{m}$ median\;

\For{$i$  $\gets$ $1$ $to$ $m$}{
$//$punish
\If{$s_{i}$ $<$ $mids*n$}{
Node $i$ is judged to be malicious and will be punished\;
}
$//$award
\If{$s_{i}$ $>$ $mids$}{
Node $i$ is judged as reward\;
$discount$=100 \% * $(m-i-1)/(m)$
}
}

\end{algorithm}

\section{Privacy Protection}

\subsection{Basic Privacy Protection}
The system relies on event notifications from smart contracts to achieve state transfer. Event notifications from smart contracts should not pass any private data. We should not use smart contracts to store private data. Smart contracts can only guarantee decentralized storage and cannot protect the privacy of data. When the submission stage and the work stage are transferred to the next stage, the problem of model exposure will be involved. Models submitted by all nodes should not be exposed to unrelated nodes in any way. So at this stage, we only use smart contracts for necessary notifications. For example, the election of committee nodes, the pledge of nodes, and the score of the final model, these parameters are the basis for non-participating nodes to participate in the system in the future. For private data, we only retain cryptographic evidence by uploading hash values. The global model obtained through aggregation is sent to participating nodes in a peer-to-peer interaction through committee nodes.

\subsection{Differential Privacy Protection}
Federated learning is also a privacy-preserving method, but recent studies have found the risk of leakage of sub-models\cite{fredrikson2015model}. For the above-mentioned basic privacy protection, privacy protection can only be achieved when all committee nodes are completely honest. In detail, each committee node will be exposed to sub-model information submitted by participating nodes, and the dishonesty of a committee will expose to all sub-models. There are many choices in the choice of privacy technology, including differential privacy\cite{abadi2016deep}, secure multi-party computation\cite{goldreich1998secure}, homomorphic encryption, and other methods. If homomorphic encryption or secure multi-party computation is used, committee members cannot judge the quality of the model. Because such algorithms inherently make the data available and invisible, this makes the committee node invisible to the quality of the submodel. This also leaves the committee with no way of judging the quality of a model.

Therefore, we apply differential privacy technology to DFLBCB. Differential privacy technology protects against model flipping attacks by adding noise to the model. The model submitted through differential privacy is a model visible to the committee, so the committee can judge the model. We believe that the committee must undertake the identification and elimination of malicious nodes. Even if it only depends on the accuracy of the model, there is no way to determine the quality of a model. However, we still believe that with the progress of theory, making the Committee nodes master more information is the key to identifying malicious nodes.
\section{Experiments and Analysis}

\subsection{Application Of Differential Privacy In DFLBCB}
We first verified the application of differential privacy and performed 150 decentralized federated learning training on the MINST dataset under the differential privacy levels of $EPS$ of 0.5, 1, 2, 4, and 8. The specific data are as follows as shown in the figure.

\centerline{\includegraphics[scale=0.6]{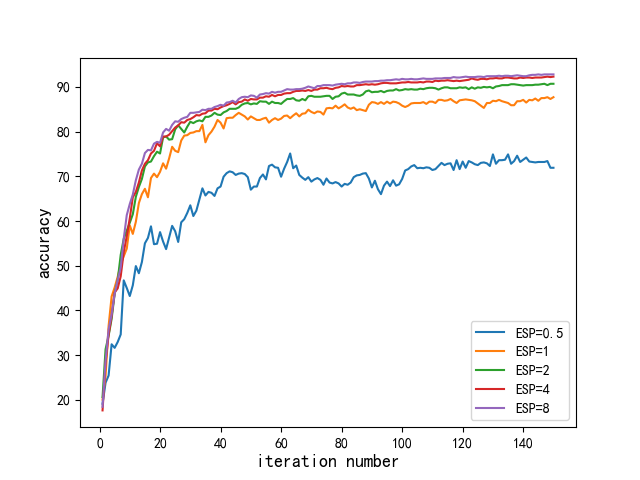}}
As shown in the figure, we verified the differential privacy method proposed in the paper \cite{abadi2016deep} in the decentralized federated learning system. The larger the $EPS$, the stronger the protection effect of differential privacy, but the availability of data also decreases, which is reflected in the decrease in the accuracy of the model. In the above experiments, when $EPS=0.5$, the availability of data decreases severely, and when $EPS=$1, 2, 4, 8, the availability of data decreases slightly, but the decrease is not large. Therefore, we recommend finding a balance between privacy protection and data availability. Different data have different effects on applying differential privacy, and it is difficult to find a general privacy protection level. We, therefore, propose to pre-test the performance of differential privacy on the dataset.

\subsection{DFLBCB Performance}

Based on the differential privacy experiments, we obtained the performance parameters of DFLBCB. The figure below shows the relationship between the number of iterations and the time-consuming.

\centerline{\includegraphics[scale=0.6]{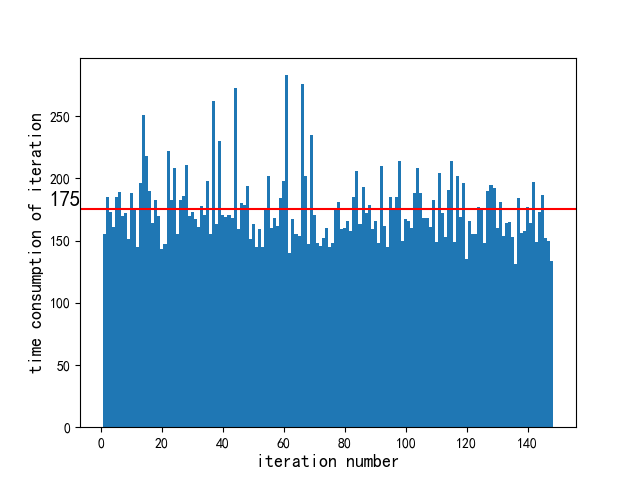}}
As shown, the average time per iteration is about 175 seconds. From the experimental results, we can see that most rounds take average time, but some rounds take significantly higher time than others. After analyzing the logs of these rounds, we found that these rounds spent a lot of time in the invocation of the state contract. Therefore, we speculate that the reason for the fluctuation is the probability of mining. Mining is essentially a probabilistic problem, and the block time is only an expected value. In the case of a few experimental rounds, there may be a delay of tens of seconds.

The figure below shows the change of the local model accuracy growth curve and the global model accuracy growth curve of a node $node$ in a committee node. 

\centerline{\includegraphics[scale=0.6]{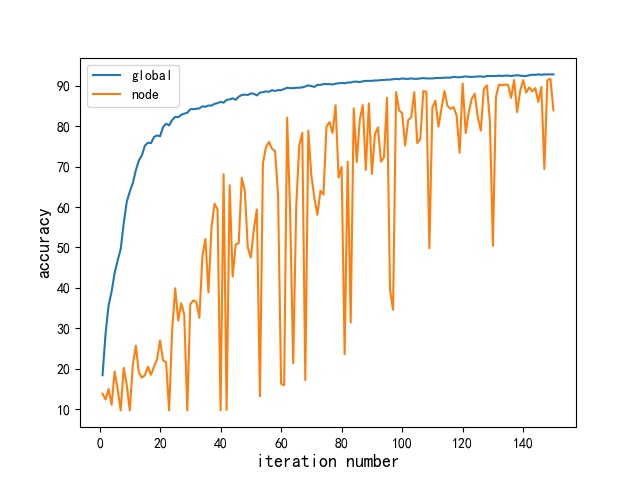}}
It is very obvious that the model of node $node$ gradually moves closer to the accuracy of the global model with the help of DFLBCB. Judging from the changing trend of the $node$ node, the node will infinitely approach the accuracy of the global model after more rounds of training.

The following figure is the proportion of time spent in the four states of DFLBCB.

\centerline{\includegraphics[scale=0.6]{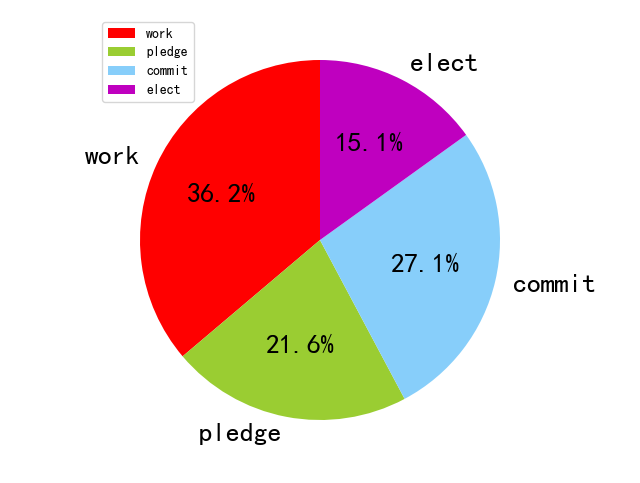}}
Among them, the $work$ stage accounts for the largest proportion, which means that in one iteration, this stage consumes the most time. In the $work$ stage, the main task of the committee node is to measure the quality of the model and aggregate sub-models, so it takes a lot of time in this stage. The time-consuming of this stage is linearly related to the size of the test set and the number of participating nodes, indicating that the larger the test set and the more participating nodes, the more time-consuming this stage will be. The second is the $model$ stage, in which the trained sub-model needs to be submitted, and the training process is time-consuming, so this stage occupies 27.1\% of the time consumption of DELBCC.

The figure below shows the time taken for a message sent by a committee node $node$ to DFLBCB and passed consensus.

\centerline{\includegraphics[scale=0.6]{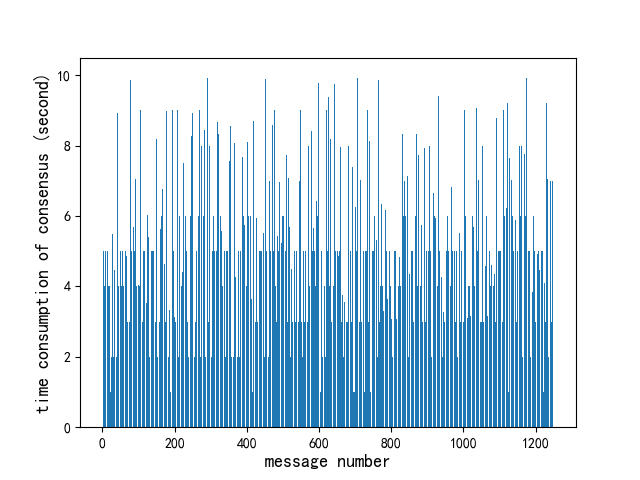}}
As shown in the figure, most of the consensus requests are completed within 5 seconds, because we set the aggregation time of requests to 5 seconds when we conducted our experiments. When a message is not aggregated in the first requested aggregation window, it must wait for the second aggregation window. In addition to the time consumption of network transmission and consensus, the time consumption of most messages is concentrated within 5 seconds, and almost all messages can complete consensus within 15 seconds.

The figure below shows the relationship between the running time of a committee node $node$ and the total amount of consensus requests sent.

\centerline{\includegraphics[scale=0.6]{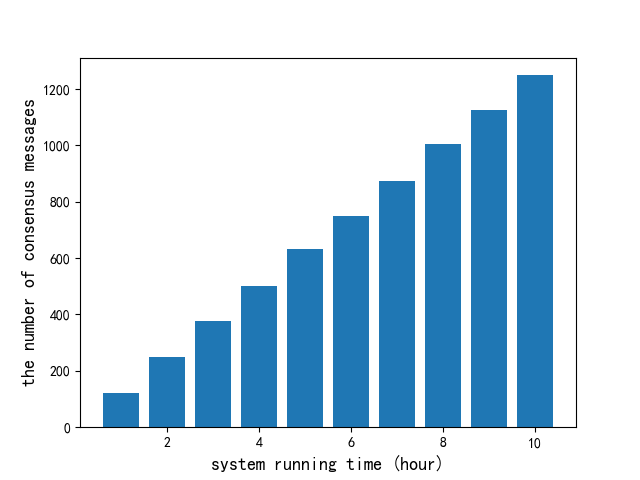}}
As shown in the figure, the hourly consensus message increment of this node is the same, indicating that the running time of each process is stable. Under the MINIST dataset, a node generates about 120 consensus data per hour. Of course, in different scenarios, due to the scale of participating nodes, the time-consuming of sub-model training and the time-consuming of sub-model testing are different, and the total amount of consensus generated by nodes per hour is also different. Or due to human intervention, the total amount of consensus generated every hour will fluctuate to varying degrees.

As shown in the figure below, we tested the performance of DFLBCB. We increment each node by 5 and test the time required for each iteration separately. To reflect the performance of DFLBCB as much as possible, we do not load any training tasks on DFLBCB. Limited by the performance of the lab server, we load up to 50 nodes on this server.

\centerline{\includegraphics[scale=0.6]{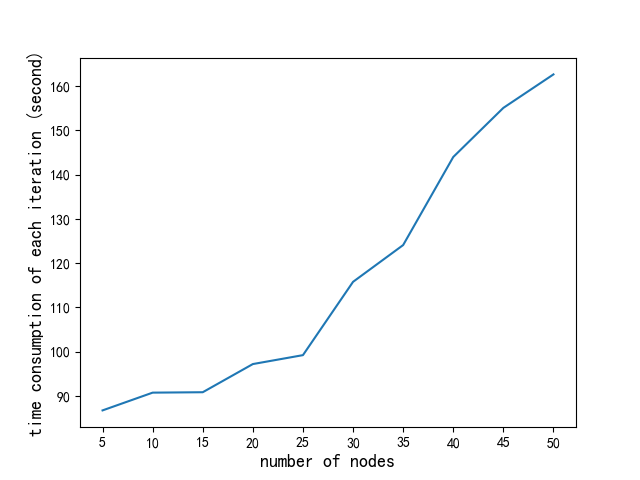}}

According to our expectation, the overall pressure of DFLBCB should be a power function with the number of nodes as a variable. The time consumption of DFLBCB is generally composed of two aspects, one is the consensus consumption, and the other is the related calculation consumption of the model. Since our experiments exclude the second consumption, all consumption of DFLBCB comes from consensus consumption. The consensus mechanism also consists of two parts, one is consumed by the PBFT algorithm, and the other is consumed by smart contract calls. Smart contract calls are constant to a certain extent. Because the consensus evidence generated by 5 nodes and 50 nodes is not significantly different in data magnitude. The other part is the time-consuming operation of the PBFT algorithm. From the behavior of the PBFT algorithm, the PBFT algorithm is time complexity of $O(n^2)$. In summary, in our experimental environment, the overall pressure of DFLBCB is a power function with the number of nodes as a variable.

From the overall experimental results, the increment of the iteration time caused by each increment of 5 nodes shows an increasing trend. In the experiments of 5-25 nodes, the consensus requests generated by each node have not put pressure on the server, so in these experiments, the time-consuming increment is relatively gentle. In experiments with 30-25 nodes, the consensus requests of each node gradually exhausted the computing resources of the server. Therefore, in these experiments, the time-consuming incremental rises faster. We predict that more nodes can be accommodated during periods where time-consuming increments are gentler if the simulations are run on higher-performance servers. Similarly, when simulation experiments are performed on lower-performance servers, fewer nodes can be accommodated during periods where the time-consuming increment is relatively flat.

\section{Future Work}
We designed and implemented a decentralized support system for federated learning. We made the following contributions: \\
1. We divide a round of federated learning processes into four stages to prevent replay attacks, use smart contracts to speed up staking and save global states, and design a state transfer scenario between the PBFT algorithm and smart contracts. \\
2. We bind the accuracy of the model to the currency and save the consensus results in advance through smart contracts. We believe that the main purpose of decentralized federated learning is to obtain models with high accuracy, so the value carrier of decentralized systems lies in the accuracy of the global model. After we bind the accuracy of the model to the value, we express the value of the model in the form of currency. This allows us to better implement the reward and punishment mechanism. \\
3. We try to use differential privacy to protect the privacy of participating nodes in a decentralized federated learning system. In order to better judge the quality of the model, we open the visibility of the committee node to the sub-model, so we do not use homomorphic encryption to protect user privacy. 

Of course, there are still many shortcomings in our design, and we will continue to solve these problems: \\
1. The model evaluation algorithm we provide is too simple and can only defend against low-level attacks, but cannot defend against well-designed attacks. According to the experimental results, when the participating nodes finally obtain the global model and train it, the accuracy of the trained model will be lower than that of the global model. For the above reasons, the accuracy of the trained model is lower than that of the untrained model. Therefore, for a node, the best practice is not to train the data but to directly submit the global model just acquired. Therefore, the evaluation algorithm of the model is still a point we need to continue to explore. \\
2. At present, we cannot solve the problem of participating nodes leaking the global model. If a node participates in training and obtains the global model, for some benefit, the node leaks the model to unrelated nodes, which causes an objective double-spending problem. In our next research, we will explore the strong binding relationship between the realization model and the currency.

\printbibliography
\nomenclature{register node}{All registered nodes.}
\nomenclature{participating nodes}{All nodes registered and participating in the current state operation of DFLBCB.}
\nomenclature{normal node}{All nodes that are registered but not participating in the current running of DFLBCB.}
\nomenclature{committee node}{Committee nodes are elected by registered nodes and are the main force for transaction processing in DFLBCB.}
\nomenclature{support node}{A node that provides the DFLBCB runtime environment.}
\nomenclature{pledge}{The process by which a registered node puts a pledge deposit into the state contract.}
\nomenclature{state contract}{A smart contract, part of a consensus system.}
\nomenclature{system status}{DFLBCB has 4 unique states, which represent different transactions currently processed by DFLBCB. They are Election, Pledge, Commit, Work.}
\nomenclature{election status}{Registered nodes can participate in the election in the election state, and ordinary nodes that are successfully elected will become committee nodes.}
\nomenclature{pledge status}{Registered nodes can decide whether to participate in this round in the pledge state. Nodes participating in this round need to pledge.}
\nomenclature{commit status}{Participating nodes that are determined to participate in this round and pledge in the pledge state must submit the locally trained sub-model.}
\nomenclature{working status}{The committee node handles submitted submodels.}
\nomenclature{accuracy-value curve}{A curve drawn by a number of points that describes the relationship between the accuracy of the model and the value.}
\nomenclature{malicious node}{A node with behavior that disrupts the normal operation of DFLBCB.}
\nomenclature{offline consensus}{Consensus on good specifications in advance at DFLBCB.}
\printnomenclature
\end{document}